\documentclass[preprintnumbers,showkeys,showpacs,byrevtex]{revtex4}
\usepackage{amsmath,amsfonts,amssymb,amscd,amsxtra,amsthm}
\usepackage{graphicx}
\usepackage{epstopdf}
\usepackage{bm}
\begin{document}      
\preprint{KIAS-P13028}
\title{Quark-gluon mixed condensate for the SU(2) light-flavor sector at finite temperature}      
\author{Seungil Nam}
\email[E-mail: ]{sinam@kias.re.kr}
\affiliation{School of Physics, Korea Institute for Advanced Study (KIAS), Seoul 130-722, Republic of Korea}
\date{\today}
\begin{abstract}
We investigate the quark-gluon mixed condensate $\langle\bar{q}\sigma\cdot Gq\rangle\equiv m^2_0\langle\bar{q}q\rangle$ for the SU(2) light-flavor sector at finite temperature ($T$). Relevant model parameters, such as the average (anti)instanton size, inter-(anti)instanton distance, and constituent-quark mass at zero virtuality, are modified as functions of $T$, employing the trivial-holonomy caloron solution. By doing that, we observe correct chiral restoration patterns depending on the current-quark mass $m$. We also perform the two-loop renormalization-group (RG) evolution for the both condensates by increasing the renormalization scale $\mu=(0.6\to2.0)$ GeV. It turns out that the mixed condensate is insensitive to the RG evolution, whereas the quark condensate become larger considerably by the evolution. Numerically, we obtain $-\langle\bar{q}\sigma\cdot Gq\rangle^{1/5}=(0.45\sim0.46)$ GeV at $T=0$ within the present theoretical framework, and the mixed condensate plays the role of the chiral order parameter for finite $T$. The ratio of the two condensates $m^2_0$ is almost flat below the chiral transition $T$ ($T_0$), and increases rapidly beyond it. From a simple linear parametrization, we obtain $m^2_0(T)/m^2_0(0)\approx (0.07,0.47)\,T/T_0+(1,0.6)$ for $(T\lesssim T_0,T\gtrsim T_0)$ at $\mu=0.6$ GeV. The present results are compared with other theoretical ones including the lattice QCD simulations, and show qualitatively good agreement with them. 
\end{abstract}
\maketitle
\section{Introduction}
Together with the active developments of the heavy-ion collision (HIC) experiments, one can now explore the hot and dense QCD matter within experimental facilities. Theoretically, the hot QCD matter can be studied appropriately from the first principle via the lattice QCD (LQCD) simulations, unlike the dense-matter LQCD, which suffers from the notorious sign problem. Beside the LQCD methods, various effective approaches have been also frequently applied to investigate the hot and/or dense QCD matter. In our previous works using the liquid instanton model at finite $T$ (LIM-$T$), we have studied various nonperturbative quantities at finite $T$ and zero quark chemical potential $\mu_q=0$, i.e. electric conductivity, shear viscosity, quark condensate, magnetic susceptibility, and so on~\cite{Nam:2013wja,Nam:2013fpa}. By comparing our numerical results with other theoretical results, including the LQCD data, we verified that LIM-$T$ provides relatively good  agreement with them. Hence, in the present work, we want to provide theoretical results for the dimension-five quark-gluon mixed condensate (MC) at finite $T$ using LIM-$T$. MC is defined by the vacuum expectation value (VEV) for the quark-gluon dimension-five operator in Minkowski space as follows:
\begin{equation}
\label{eq:QDEF}
\langle\bar{q}\sigma\cdot Gq\rangle\equiv m^2_0\langle\bar{q}q\rangle,
\end{equation}
where $q$ and $G_{\mu\nu}\equiv G^a_{\mu\nu}\lambda^a/2$ stands for the quark and gluon fields. The antisymmetric tensor is given as $\sigma_{\mu\nu}=i(\gamma_\mu\gamma_\nu-\gamma_\nu\gamma_\mu)/2$. As shown in  Eq.~(\ref{eq:QDEF}), MC is usually factorized into a mass constant $m_0$ and the quark condensate $\langle\bar{q}q\rangle$ (QC) for a practical usage in the operator-product expansion (OPE) of the QCD sum rule (QCDSR) calculations. In other words, $m^2_0$ in the right-hand-side of Eq.~(\ref{eq:QDEF}) plays an important input value in applying the QCDSR technique. For vacuum $(T,\mu_q)=0$,  many theoretical works have been done so far for MC using LQCD~\cite{Doi:2002pq}, holographic QCD (hQCD)~\cite{Kim:2008ff}, global color-symmetry model~\cite{Zong:2002mh}, Dyson-Schwinger method~\cite{Lu:2010zze}, LIM~\cite{Polyakov:1996kh,Nam:2006ng}, hybrid-current method~\cite{Aladashvili:1995zj}, truncated quark-quark interaction model~\cite{Meissner:1997ks}, and so on. Note that, however, there have been only a few theoretical studies for MC at finite $T$ so far: LQCD~\cite{Doi:2004xe} and effective approach, such as the global color-symmetry model~\cite{Zhang:2004xg}. Moreover, the LQCD work was done via the Kogut-Susskind (KS) fermion at the quenched level, giving $T_c=280$ MeV, which is much higher than those from the recent full LQCD simulations~\cite{Doi:2004xe}. Hence, considering the recent progress in the realistic LQCD simulations at finite $T$, it is worth to providing sophisticated theoretical estimations for MC, as done in the previous works, and the numerical results will be a useful guide for future LQCD simulations. 

Here, we want to make a brief description for LIM-$T$ as a theoretical framework: LIM-$T$ is basically based on LIM, which manifests the nonlocal quark-quark interactions via the fermionic zero mode. To extend LIM to a finite-$T$ system, the model parameters of LIM, i.e. the average (anti)instanton size $(\bar{\rho})$ and inter-(anti)instanton distance $\bar{R}$, are modified as functions of $T$ using the trivial-holonomy caloron solution~\cite{Harrington:1976dj,Diakonov:1988my,Nam:2009nn}. Note that these modifications result in the partial chiral restoration of the parameters. Constructing an effective thermodynamic potential within the same model, we compute the constituent-quark mass at zero virtuality as a function of $T$, as an chiral order parameter, by solving the saddle-point equation with respect to the variation parameter~\cite{Diakonov:2002fq}. As a result, one observe the second-order and crossover chiral restoration patterns for the zero and finite current-quark masses ($m$), satisfying the universal restoration patterns. We note that the chiral transition $T$ is determined as $T_0\approx151$ MeV for $m=(0,5)$ MeV. Along with these $T$-modified parameters and constituent-quark mass, we employ the fermionic Matsubara formula for the relevant matrix elements. We also perform the two-loop renormalization-group (RG) evolution for the both condensates, MC and QC, for $\mu=(0.6\to2.0)$ GeV, to compare the numerical results with the LQCD data. Note that $\mu=0.6$ GeV is a typical scale of LIM, i.e. $\mu\approx1/\bar{\rho}$. From the numerical results, it turns out that MC is insensitive to the evolution, whereas QC gets larger by the evolution. Numerically, we obtain $-\langle\bar{q}\sigma\cdot Gq\rangle^{1/5}=(0.45\sim0.46)$ GeV at $T=0$ within the present theoretical framework. Moreover, MC as well as QC play the role of the order parameters of the chiral phase transition. The ratio of the two condensates $m^2_0$ turns out to be almost flat below the chiral transition $T$ ($T_0$), and increases stiffly beyond it. From the numerical results, we parametrize the ratio approximately with a simple linear functions at $\mu=0.6$ GeV:
\begin{equation}
\label{eq:PARAMSQR4}
m^2_0(T)/m^2_0(0)\approx (0.07,0.47)\,T/T_0+(1,0.6)\,\,\,\,\mathrm{for}\,\,\,\,(T\lesssim T_0,T\gtrsim T_0).
\end{equation}

We organize the present work as follows: In Section II, we briefly introduce the liquid instanton model (LIM) and how to compute the mixed condensate in terms of the field theoretical manner. In Section III, the temperature modifications of the relevant model parameters are performed using the trivial caloron solution. We also show the correct universal chiral restoration patterns, computed within the present model.  The numerical results for MC, QC, and their ratio $m^2_0$ as functions of temperature are presented with relevant discussions in Section IV. Final Section is devoted to summary, conclusion, and future perspectives.

\section{Theoretical framework}
\begin{figure}[t]
\includegraphics[width=14cm]{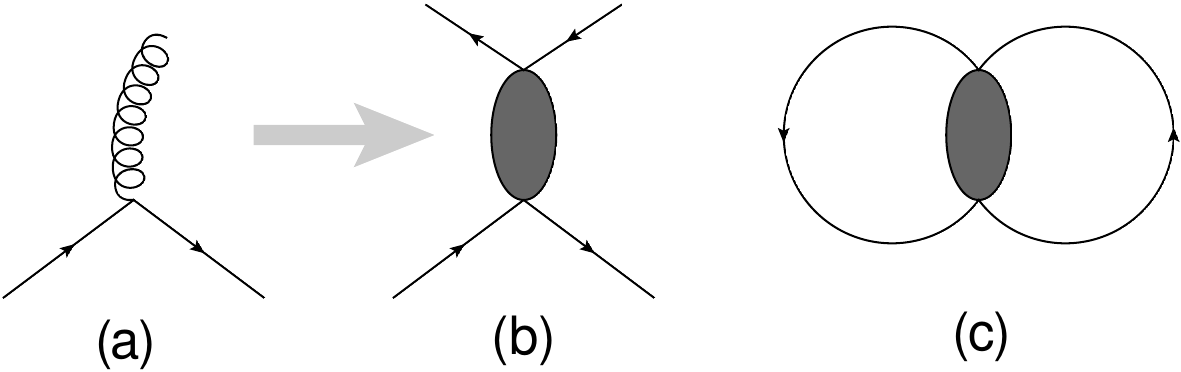}
\caption{(a) The quark-gluon vertex given with the quark (solid) and gluon (wiggle) fields. (b) The gluon field is effectively replaced by the quark-instanton vertex on the right, i.e. effective quark-gluon vertex $Y_{\pm,1}(x,U)$ in Eq.~(\ref{eq:Y}). (c) The quark-gluon mixed condensate in Eq.~(\ref{eq:MC0}), constructed by attaching the external quark lines in the effective quark-gluon vertex (b).}       
\label{FIG1}
\end{figure}

In this Section, we elucidate the theoretical framework and how to compute the relevant condensates. All the calculations will be performed in the leading large-$N_c$ limit. We make use of an effective action, derived from the liquid-instanton model (LIM) in Euclidean space for the SU(2) light-flavor sector, and it reads
\begin{equation}
\label{eq:EA}
\mathcal{S}_\mathrm{eff}=-\mathrm{Sp}
\ln\left[i\rlap{/}{\partial}+i\hat{m}+iM_0F^2(\partial) \right],
\end{equation}
where the current-quark mass matrix for SU($2_f$) is designated as $\hat{m}=\mathrm{diag}(m_u,m_d)$. By assuming the isospin symmetry, we set $m_u\approx m_d=m=5$ MeV throughout the present work, when the current-quark mass is finite. $M_0$ denotes the effective quark mass at zero virtuality, i.e. constituent-quark mass. In the Euclidean momentum space, the quark momentum distribution $F(k)$ is given by
\begin{equation}
\label{eq:FK}
F(k)=2\tau
\left[I_0(\tau)K_1(\tau)i_1(\tau)K_0(\tau)-\frac{1}{\tau}I_1(\tau)K_1(\tau) \right],
\,\,\,\,\tau=\frac{|k|\bar{\rho}}{2}.
\end{equation}
Here, $I_n$ and $K_n$ indicate the modified Bessel functions, whereas $\bar{\rho}$ stands for the average (anti)instanton size, which corresponds to the inverse of the model scale, $\mu\approx1/\bar{\rho}$. First, we want to calculate $\langle\bar{q}q\rangle$ (QC) by functional differentiating $\mathcal{S}_\mathrm{eff}$ with respect to $m$, resulting in
\begin{equation}
\label{eq:CC}
-\langle\bar{q}q\rangle=\frac{1}{N_f}\frac{\delta\mathcal{S}_\mathrm{eff}}{\delta m}=4N_c\int\frac{d^4k}{(2\pi)^4}\left[\frac{\bar{M}_k}{k^2+\bar{M}^2_k}-\frac{m}{k^2+m^2}\right],
\end{equation}
where we have used a simplified notation $\bar{M}_k=m+M(k)=m+M_k$, and $N_c$ indicates the number of color: $N_c=3$. Note that $\bar{\rho}$ and $\bar{R}$ at $T=0$ were estimated by the variational method $(\bar{\rho},\bar{R})\approx(0.35,0.95)$ fm~\cite{Diakonov:2002fq}, phenomenological way $(\bar{\rho},\bar{R})\approx(1/3,1)$ fm~\cite{Shuryak:1981ff}, and the LQCD simulation $(\bar{\rho},\bar{R})\approx(0.36,0.89)$ fm~\cite{Chu:1994vi}. Among them, we employ the phenomenological values, considering about $10\%$ uncertainties within those estimations. The value for $M_0$ for vacuum can be fixed by the self-consistent equation:
\begin{equation}
\label{eq:SCE}
\frac{N}{V}=\frac{1}{\bar{R}^4}=4N_c\int\frac{d^4k}{(2\pi)^4}\frac{M^2_k}{k^2+M^2_k},
\end{equation}
where $N/V$ indicates the instanton number density or packing fraction of the instanton ensemble. By solving Eq.~(\ref{eq:SCE}) with the phenomenological instanton parameters given above, one acquires $M_0\approx350$ MeV. 

Now, we are in a position to discuss the quark-gluon operator in terms of the quark fields. The local operator inside MC in the left-hand-side of Eq.~(\ref{eq:QDEF}) corresponds to the quark-gluon interaction of Yukawa type, as shown in (a) of Figure~\ref{FIG1}. However, in this model calculation, the gluon field strength ($G_{\mu\nu}$) can be expressed in terms of the nontrivial quark-instanton interaction~\cite{Polyakov:1996kh}. First, the one flavor quark and one instanton interaction can be written as a function of $x$ and color orientation matrix $U$.  
\begin{equation}
\label{eq:Y}
Y_{\pm,1}(x,U)=(2\pi\bar{\rho})^2\int\frac{d^4k}{(2\pi)^4}\frac{d^4p}{(2\pi)^4}
 F(k\bar{\rho}) F(p\bar{\rho})e^{ix\cdot(k-p)}\left[U^{\alpha}_{i'}(U^{j'}_{\beta})^{\dagger}\epsilon^{ii'}\epsilon_{jj'}\right]\left[iq^{\dagger}(k)_{\alpha i}\frac{1\pm\gamma_5}{2}q(p)^{\beta j}\right].
\end{equation}
Here, $\alpha$ and $i$ for $q_{\alpha i}$denote the color and spinor indices, respectively. In deriving Eq.~(\ref{eq:Y}), we assumed the $\delta$-function type instanton distribution $\sim\delta(\rho-\bar{\rho})$ as usual. A schematic diagram for Eq.~(\ref{eq:Y}) is given in (b) of Figure~\ref{FIG1}. Then, we write the field strength tensor $F^a_{\mu\nu}$ in terms of instanton configuration:
\begin{equation}
\label{eq:FA}
F^a_{\pm\mu\nu}(x,x',U)
=\frac{1}{2}\left[\lambda^aU\lambda^bU^{\dagger}\right]F^b_{\pm\mu\nu}(x'-x).
\end{equation}
$F^b_{\pm\mu\nu}(x'-x)$ stand for the field strength consisted of a certain
instanton configuration.  Using Eqs.~(\ref{eq:Y}) and (\ref{eq:FA}), we can
define the field strength tensor in momentum space in terms of the quark-instanton interactions:
\begin{equation}
\label{eq:FFF}
\hat{F}^a_{\pm\mu\nu}=\frac{iN_cM}{4\pi\bar{\rho}^2}\int d^4x\int dUF^a_{\pm\mu\nu}(x,x',U)Y_{\pm,1}(x,U) 
\end{equation}
Following the course of Ref.~\cite{Polyakov:1996kh}, finally, we obtain the
mixed-condensate as follows:
\begin{equation}
\label{eq:MC0}
\langle\bar{q}\sigma\cdot Gq\rangle=2N_c\int\frac{d^4k}{(2\pi)^4}
\int\frac{d^4p}{(2\pi)^4}\frac{\sqrt{M_kM_p}G_{k,p}N_{k,p}}
{[k^2+\bar{M}^2_k][p^2+\bar{M}^2_p]},
\end{equation}
where $G_{k,p}$, which relates to the Fourier transform of the instanton field
strength, and $N_{k,p}$ are defined by the followings:
\begin{eqnarray}
\label{eq:GN}
G_{k,p}&=&32\pi^2\bar{\rho}^2\left[\frac{K_0(t)}{2}+\frac{4K_0(t)}{t^2}
+\left(\frac{2}{t}+\frac{8}{t^3}\right)K_1(t)-\frac{8}{t^4}\right],
\,\,\,\,t=|k-p|\bar{\rho}
\cr
N_{k,p}&=&\frac{1}{4}\mathrm{Tr}_\gamma
\left[\sigma_{\mu\nu}(\rlap{/}{k}+i\bar{M}_k)\Gamma_{\mu\nu}(\rlap{/}{p}+i\bar{M}_p) \right],\,\,\,\,
\Gamma_{\mu\nu}=\sigma_{\alpha\nu}\frac{q_\rho q_\mu}{q^2}
+\sigma_{\mu\alpha}\frac{q_\rho q_\nu}{q^2}-\frac{1}{2}\sigma_{\mu\nu},\,\,\,\,
q=k-p.
\end{eqnarray}
Performing the trace of the Lorentz index and simplifying the expression, we arrive at
\begin{equation}
\label{eq:MCB}
\langle\bar{q}\sigma\cdot Gq\rangle
=4N_c\int\frac{d^4k}{(2\pi)^4}\frac{d^4p}{(2\pi)^4}
\frac{\sqrt{M_kM_p}G_{k,p}\left[(k\cdot p)-\frac{4(k\cdot q)(p\cdot q)}{q^2} \right]}
{[k^2+\bar{M}^2_k][p^2+\bar{M}^2_p]},
\end{equation}
which is schematically represented by (c) in Figure~\ref{FIG1}.

MC in Eq.~(\ref{eq:MCB}) can be written as a function of $T$, employing the fermionic Matsubara formula. Note that one of the technical difficulties here is the double summation over the Matsubara frequencies $\omega_n=(2n+1)\pi T$ and $\omega_{n'}=(2n'+1)\pi T$, corresponding to the integral over $k_4$ and $p_4$. However, this difficulty can be easily removed by assuming that $(k_4,p_4)\to0$ in the numerator in Eq.~(\ref{eq:MCB}), whereas we rewrite the denominator by $(k_4,p_4)\to(\omega_n,\omega_{n'})$. According to this simplification, one can separate the integrand in Eq.~(\ref{eq:MCB}) into independent summations over $\omega_n$ and $\omega_{n'}$, resulting in
\begin{eqnarray}
\label{eq:MCBT}
\langle\bar{q}\sigma\cdot Gq\rangle
&\approx&4N_cT^2\sum_{n}\sum_{n'}
\int\frac{d^3\bm{k}}{(2\pi)^3}\frac{d^3\bm{p}}{(2\pi)^3}
\frac{\sqrt{M_{\bm{k}}M_{\bm{p}}}G_{\bm{k},\bm{p}}
\left[(\bm{k}\cdot \bm{p})-\frac{4(\bm{k}\cdot \bm{q})(\bm{p}\cdot \bm{q})}{\bm{q}^2} \right]}
{[\omega^2_n+E^2_{\bm{k}}][\omega^2_{n'}+E^2_{\bm{p}}]}
\cr
&=&N_c\int\frac{d^3\bm{k}}{(2\pi)^3}\frac{d^3\bm{p}}{(2\pi)^3}
\frac{\sqrt{M_{\bm{k}}M_{\bm{p}}}G_{\bm{k},\bm{p}}}{E_{\bm{k}}E_{\bm{p}}}\left[(\bm{k}\cdot \bm{p})-\frac{4(\bm{k}\cdot \bm{q})(\bm{p}\cdot \bm{q})}{\bm{q}^2} \right]
\mathrm{tanh}\left[\frac{E_{\bm{k}}}{2T} \right]
\mathrm{tanh}\left[\frac{E_{\bm{p}}}{2T} \right],
\end{eqnarray}
where we have used $E^2_{\bm{k}}=\bm{k}^2+\bar{M}^2_{\bm{k}}$ for brevity. Note that $M_{\bm{k}}$ and $G_{\bm{k},\bm{p}}$ in the above equation represent those in Eqs.~(\ref{eq:FK}) and (\ref{eq:GN}) by replacing $(k,p)\to(\bm{k},\bm{p})$ as
\begin{equation}
\label{eq:MASSSS}
M_k\to M_{\bm{k}}=M_0\left[\frac{2}{2+\bm{k}^2\bar{\rho}^2} \right]^3,\,\,\,\,
G_{k,p}(\bar{\rho})\to G_{\bm{k},\bm{p}}(\bar{\rho})
\end{equation}
to reproduce the correct low-energy constants~\cite{sinam}. Note that $M_{\bm{k}}$ in Eq.~(\ref{eq:MASSSS}) show stronger decrease with respect to $|\bm{k}|$, in comparison to those used in the previous works~\cite{Nam:2013wja,Nam:2013fpa}. The reason to use this stronger one is to reproduce the chiral transition $T$, $T_0=(154\pm9)$ MeV for $m\ne0$, which was estimated by the recent LQCD simulations, using the $2+1$ flavor QCD thermodynamics with improved staggered fermions~\cite{Bazavov:2011nk,Bazavov:2012ty}. Detailed discussions on $T_0$ will be given in Section IV.

\section{Temperature-dependent model parameters}
Now, we want to address how to determine the $T$-dependent effective quark mass $M_0$ in Eq.~(\ref{eq:MASSSS}). In Refs.~\cite{Nam:2009nn}, we derived it by using the caloron distribution with the trivial holonomy, i.e. Harrington-Shepard caloron~\cite{Harrington:1976dj,Diakonov:1988my}. Firstly, we want to explain briefly how to modify $\bar{\rho}$ and $\bar{R}$ as functions of $T$, using the caloron solution. Details can be found in Ref.~\cite{Nam:2009nn}.  An instanton distribution function for arbitrary $N_c$ and $N_f$ can be written with a Gaussian suppression factor as a function of $T$ and an arbitrary instanton size $\rho$ for pure-glue QCD~\cite{Diakonov:1988my}:
\begin{equation}
\label{eq:IND}
d(\rho,T)=\underbrace{C_{N_c}\,\Lambda^b_{\mathrm{RS}}\,
\hat{\beta}^{N_c}}_\mathcal{C}\,\rho^{b-5}
\exp\left[-(A_{N_c}T^2
+\bar{\beta}\gamma n\bar{\rho}^2)\rho^2 \right].
\end{equation}
We note that the CP-invariant vacuum was taken into account in Eq.~(\ref{eq:IND}), and we assumed the same analytical form of the distribution function for the (anti)instanton. Note that the instanton number density (packing fraction) $N/V\equiv n\equiv1/\bar{R}^4$ and $\bar{\rho}$ have been taken into account as functions of $T$ implicitly. For simplicity, we take the numbers of the anti-instanton and instanton are the same, i.e. $N_I=N_{\bar{I}}=N$. We also assigned the constant factor in the right-hand-side of the above equation as $\mathcal{C}$ for simplicity. The abbreviated notations read:
\begin{eqnarray}
\label{eq:PARA}
\hat{\beta}&=&-b\ln[\Lambda_\mathrm{RS}\rho_\mathrm{cut}],\,\,\,\,
\bar{\beta}=-b\ln[\Lambda_\mathrm{RS}\langle R\rangle],\,\,\,
C_{N_c}=\frac{4.60\,e^{-1.68\alpha_{\mathrm{RS}} Nc}}{\pi^2(N_c-2)!(N_c-1)!},
\cr
A_{N_c}&=&\frac{1}{3}\left[\frac{11}{6}N_c-1\right]\pi^2,\,\,\,\,
\gamma=\frac{27}{4}\left[\frac{N_c}{N^2_c-1}\right]\pi^2,\,\,\,\,
b=\frac{11N_c-2N_f}{3}.
\end{eqnarray}
Note that we defined the one-loop inverse charges $\hat{\beta}$ and $\bar{\beta}$ at certain phenomenological cutoffs $\rho_\mathrm{cut}$ and $\langle R\rangle\approx\bar{R}$. $\Lambda_{\mathrm{RS}}$ denotes a scale, depending on a renormalization scheme, whereas $V_3$ for the three-dimensional volume. Using the instanton distribution function in Eq.~(\ref{eq:IND}), we can compute the average value of the instanton size $\bar{\rho}^2$ straightforwardly as follows~\cite{Schafer:1996wv}:
\begin{equation}
\label{eq:rho}
\bar{\rho}^2(T)
=\frac{\int d\rho\,\rho^2 d(\rho,T)}{\int d\rho\,d(\rho,T)}
=\frac{\left[A^2_{N_c}T^4
+4\nu\bar{\beta}\gamma n \right]^{\frac{1}{2}}
-A_{N_c}T^2}{2\bar{\beta}\gamma n},
\end{equation}
where $\nu=(b-4)/2$. It can be easily shown that Eq.~(\ref{eq:rho}) satisfies the  following asymptotic behaviors~\cite{Schafer:1996wv}:
\begin{equation}
\label{eq:asym}
\lim_{T\to0}\bar{\rho}^2(T)=\sqrt{\frac{\nu}{\bar{\beta}\gamma n}},
\,\,\,\,
\lim_{T\to\infty}\bar{\rho}^2(T)=\frac{\nu}{A_{N_c}T^2}.
\end{equation}
Here, the second relation of Eq.~(\ref{eq:asym}) indicates a correct scale-temperature behavior at high $T$, i.e., $1/\bar{\rho}\approx\Lambda\propto T$. Substituting Eq.~(\ref{eq:rho}) into Eq.~(\ref{eq:IND}), the caloron distribution function can be evaluated further:
\begin{equation}
\label{eq:dT}
d(\rho,T)=\mathcal{C}\,\rho^{b-5}
\exp\left[-\mathcal{F}(T)\rho^2 \right],\,\,\,\,
\mathcal{F}(T)=\frac{1}{2}A_{N_c}T^2+\left[\frac{1}{4}A^2_{N_c}T^4
+\nu\bar{\beta}\gamma n \right]^{\frac{1}{2}}.
\end{equation}
The instanton packing fraction $n$ can be computed self-consistently, using the following equation:
\begin{equation}
\label{eq:NOVV}
n^\frac{1}{\nu}\mathcal{F}(T)=\left[\mathcal{C}\,\Gamma(\nu) \right]^\frac{1}{\nu},
\end{equation}
where we replaced $NT/V_3\to n$, and $\Gamma(\nu)$ stands for the $\Gamma$-function with an argument $\nu$. Note that $\mathcal{C}$ and $\bar{\beta}$ can be determined easily using Eqs.~(\ref{eq:rho}) and (\ref{eq:NOVV}), incorporating the vacuum values for $n\approx(200\,\mathrm{MeV})^4$ and $\bar{\rho}\approx(600\,\mathrm{MeV})^{-1}$: $\mathcal{C}\approx9.81\times10^{-4}$ and $\bar{\beta}\approx9.19$. Finally, in order for estimating the $T$-dependence of $M_0$, it is necessary to consider the normalized distribution function, defined as follows,
\begin{equation}
\label{eq:NID}
d_N(\rho,T)=\frac{d(\rho,T)}{\int d\rho\,d(\rho,T)}
=\frac{\rho^{b-5}\mathcal{F}^\nu(T)
\exp\left[-\mathcal{F}(T)\rho^2 \right]}{\Gamma(\nu)}.
\end{equation}
Here, the subscript $N$ denotes the normalized distribution. For brevity, we want to employ the large-$N_c$ limit to simplify the expression for $d_N(\rho,T)$. In this limit, as understood from Eq.~(\ref{eq:NID}), $d_N(\rho,T)$ can be approximated as a $\delta$-function:
\begin{equation}
\label{eq:NID2}
\lim_{N_c\to\infty}d_N(\rho,T)=\delta[{\rho-\bar{\rho}(T)}].
\end{equation}

The numerical result for $\bar{\rho}(T)$ is given in the left panel of Figure~\ref{FIG23}. The curve for $\bar{\rho}(T)$ shows that the average (anti)instanton size smoothly decreases with respect to temperature. This behavior  indicates that the instanton ensemble gets diluted and the nonperturbative effects via the quark-instanton interactions are diminished. At $T=(150\sim200)$ MeV, which is close to the chiral phase transition temperature, the instanton size decreases by about $(10\sim20)\%$ in comparison to its value at $T$=0. Considering that the instanton size corresponds to the scale parameter of the model, i.e. UV cutoff mass, $\bar{\rho}\approx1/\Lambda$, the temperature-dependent cutoff mass is a clearly distinctive feature in comparison to other low-energy effective models, such as the NJL model. In addition, we also show the temperature dependence of the average (anti)instanton number density or (anti)instanton packing fraction, $N/V$, in the left panel of Figure~\ref{FIG23}. Again, the instanton number density decreases as temperature increases: The instanton ensemble becomes diluted with respect to $T$. 

\section{Thermodynamics potential and saddle-point equation}
\begin{figure}[t]
\begin{tabular}{cc}
\includegraphics[width=8.5cm]{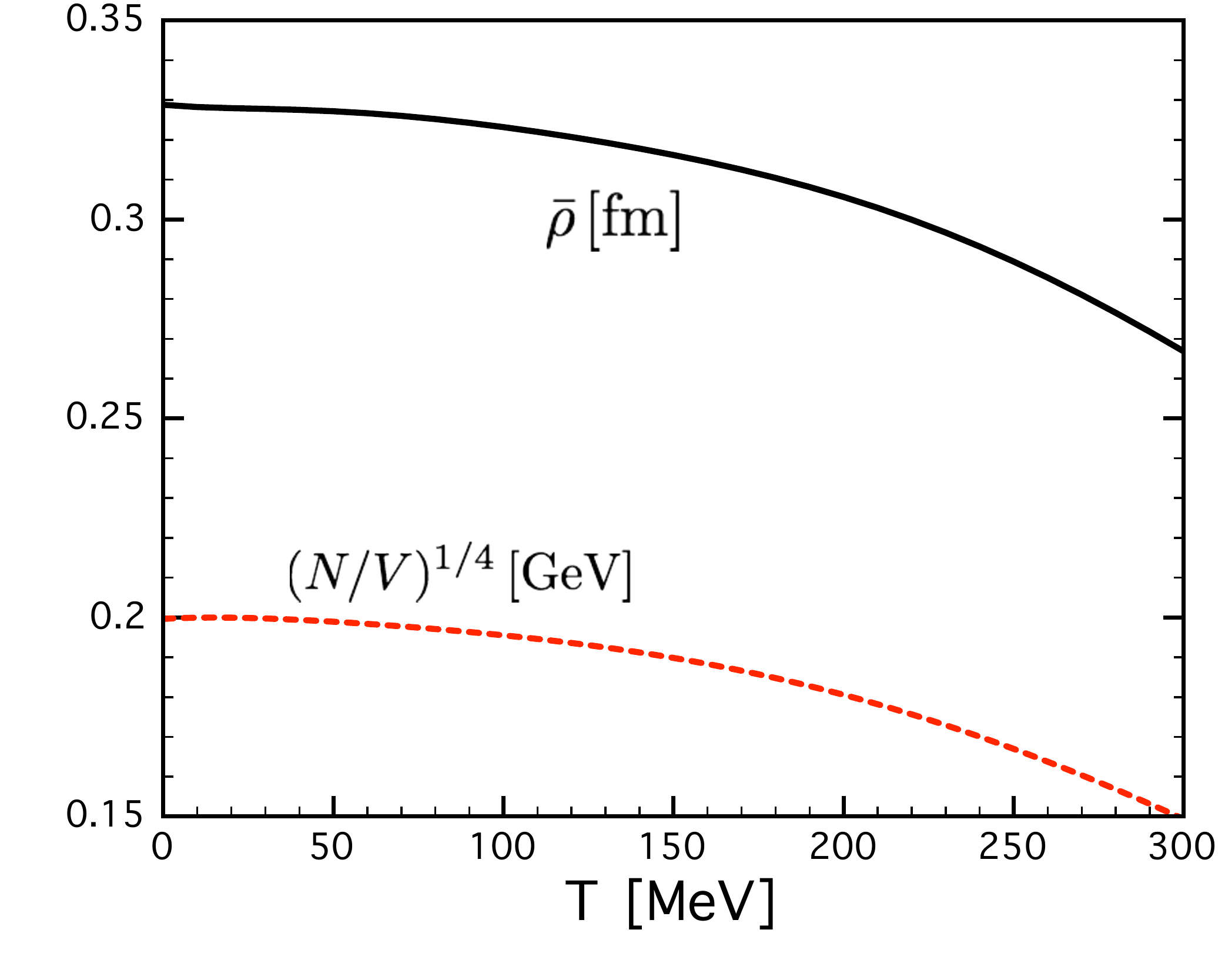}
\includegraphics[width=8.5cm]{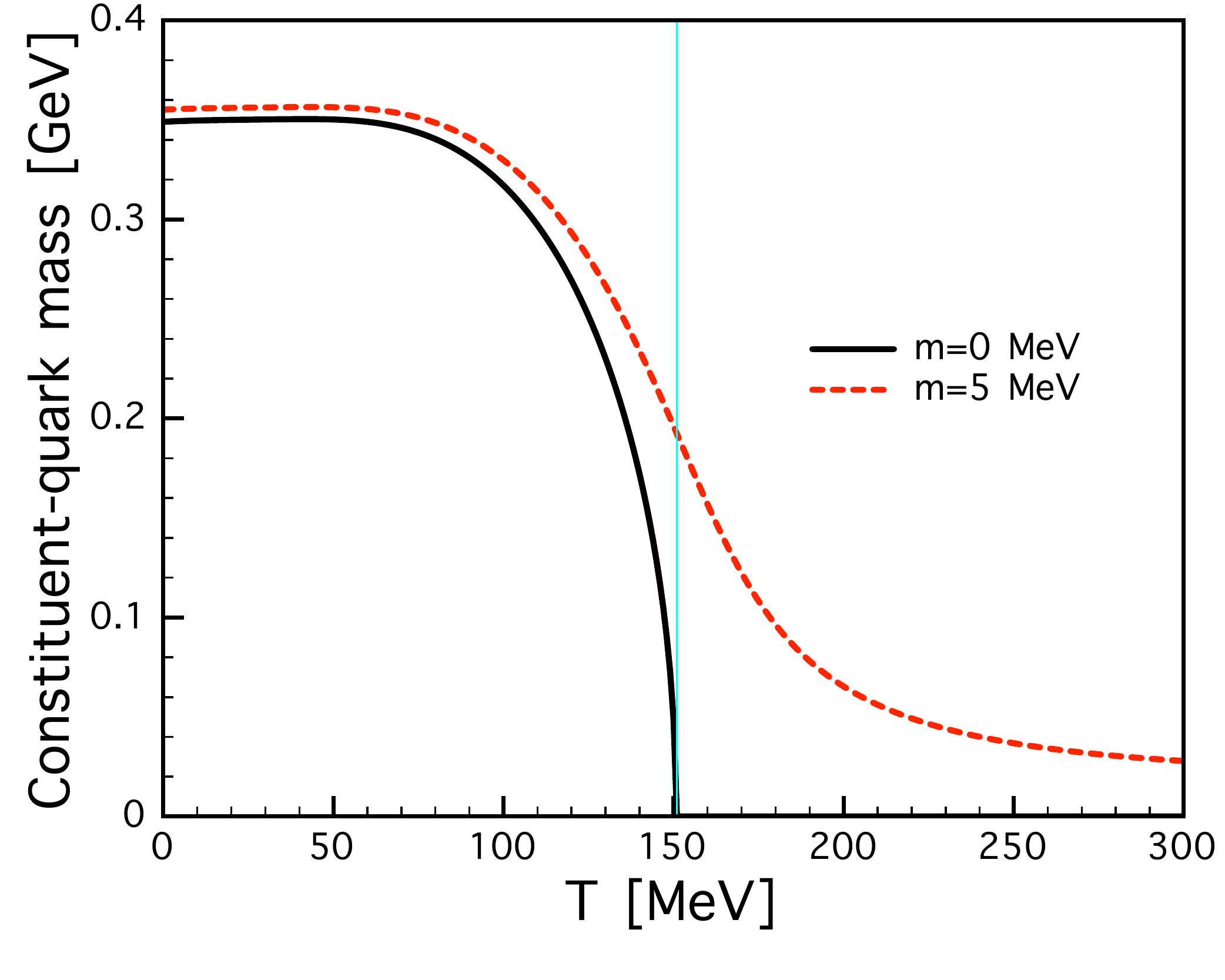}
\end{tabular}
\caption{(Color online) Left: LIM-$T$ parameters $\bar{\rho}$ [fm] and $1/\bar{R}=(N/V)^{1/4}$ [GeV] as functions of $T$ from the caloron distribution for $N_c=3$, as discussed in Section III. Right: Constituent-quark mass $M_0$ [GeV] for $N_c=3$ as a function of $T$ for $m=(0,5)$ MeV, given in the solid and dash lines, respectively. The vertical line indicates the chiral phase transition temperature $T_0\approx151$ MeV.}       
\label{FIG23}
\end{figure}

As in Ref.~\cite{Nam:2009nn}, the LIM-$T$ thermodynamic potential per volume in the leading $1/N_c$ contributions at zero quark chemical potential can be written as follows:
\begin{eqnarray}
\label{eq:TP}
\Omega_\mathrm{LIM}
&=&\frac{N}{V}\left[1-\ln\frac{N}{\lambda V\mathrm{M}} \right]+2\sigma^2-2N_cN_f\int^\infty_0\frac{d^3\bm{k}}{(2\pi)^3}
\left[E_{\bm{k}}+2T\ln\left[1+e^{-\frac{E_{\bm{k}}}{T}}  \right]
 \right],
\end{eqnarray}
where $\lambda$ and $\mathrm{M}$ represent a Lagrange multiplier to exponentiate the effective quark-instanton action and an arbitrary mass parameter to make the argument for the logarithm dimensionless. $\sigma$ stands for the isosinglet scalar meson field corresponding to the effective quark mass. In the leading large-$N_c$ contributions, we have the relation $2\sigma^2=N/V$~\cite{Nam:2009nn}. Then, the saddle-point equation can be derived from Eq.~(\ref{eq:TP}) by differentiating $\Omega_\mathrm{LIM}$ by the Lagrange multiplier $\lambda$:
\begin{equation}
\label{eq:LIMGAP}
\frac{\partial\Omega_\mathrm{LIM}}{\partial \lambda}=0\to
\frac{N_f}{\bar{M}_0}\frac{N}{V}-2N_cN_f\int^\infty_0\frac{d^3\bm{k}}{(2\pi)^3}
\frac{M_{\bm{k}}}{E_{\bm{k}}}\left[1-\frac{2e^{-\frac{E_{\bm{k}}}{T}}}{1+e^{-\frac{E_{\bm{k}}}{T}}}\right]=0.
\end{equation}
Here, $\bar{\rho}=\bar{\rho}(T)$ and $M_0=M_0(T)$ implicitly. Note that one can write the instanton number density in terms of the effective quark mass $M_0$ and $\bar{\rho}$~\cite{Diakonov:2002fq}:
\begin{equation}
\label{eq:NOV}
\frac{N}{V}=\frac{\mathcal{C}_0N_cM^2_0}{\pi^2\bar{\rho}^2}.
\end{equation}
The value of the real-positive parameter $\mathcal{C}_0$ is determined to reproduce $M_0=(350,355)$ MeV for $m=(0,5)$ at $T=0$, resulting in $\mathcal{C}_0=(0.434,0.438)$, respectively, by solving the saddle-point equation in Eq.~(\ref{eq:LIMGAP}). By solving Eq.~(\ref{eq:LIMGAP}) with respect to $M_0$ numerically, the numerical results for $M_0$ as a function of $T$ are given in the right panel of Figure~\ref{FIG23} for the zero and finite current quark mass: $m=0$ (solid) and $m=5$ MeV (dotted). These results show proper universal patterns for the chiral phase transition, i.e. the second-order chiral phase transition for the massless quark and the crossover for the finite mass.  From those numerical results, the phase transition temperatures are given as $T_0\approx151$ MeV for $m=(0,5)$ MeV. $T_0$ is marked by the vertical line in the right panel of Figure~\ref{FIG23}. Note that $T_0$ for the second-order and crossover chiral phase transition are almost consistent to each other. The obtained $T_0$ is compatible with the LQCD estimation as mentioned above~\cite{Bazavov:2011nk,Bazavov:2012ty}.

\section{Numerical results and discussions}
In this Section, we provide the numerical results for MC, QC, and their ratio MC/QC$\equiv m^2_0$ in Eq.~(\ref{eq:QDEF}) as functions of $T$ for different current-quark masses and renormalization scales. First, we want to explain briefly how to perform the RG evolution for those condensates. The RG evolution can be performed via the following equations for them, computed at different renormalization scales $\mu_{i,f}$:
\begin{equation}
\label{eq:RGG}
\langle\bar{q}\sigma\cdot G q\rangle_{\mu_f}=\left[\frac{\alpha_s(\mu_i)}{\alpha_s(\mu_f)} \right]^{\frac{\gamma_\mathrm{MC}}{b}}
\langle\bar{q}\sigma\cdot G q\rangle_{\mu_i},\,\,\,\,
\langle\bar{q}q\rangle_{\mu_f}=\left[\frac{\alpha_s(\mu_i)}{\alpha_s(\mu_f)} \right]^{\frac{\gamma_\mathrm{QC}}{b}}
\langle\bar{q} q\rangle_{\mu_i},
\end{equation}
where $b=11N_c/3-2N_f/3=29/3$, and the anomalous dimensions for QC and MC are given by $\gamma_\mathrm{MC}=-2/3$ and $\gamma_\mathrm{QC}=4$. Performing the two-loop RG evolution for $\mu_i\to\mu_f$~\cite{MUTA}, we obtain the following value for the ratio of the strong couplings:
\begin{equation}
\label{eq:RG}
\mathcal{R}(\mu_i\to \mu_f)\equiv\frac{\alpha_s(\mu_i)}{\alpha_s(\mu_f)}\approx 
\frac{\ln[\mu_f/\Lambda_\mathrm{QCD}]\left[1-\frac{\beta_1}{\beta^2_0}
\frac{\ln[\ln[\mu^2_i/\Lambda^2_\mathrm{QCD}]]}{\ln[\mu^2_i/\Lambda^2_\mathrm{QCD}]} \right]}
{\ln[\mu_i/\Lambda_\mathrm{QCD}]\left[1-\frac{\beta_1}{\beta^2_0}
\frac{\ln[\ln[\mu^2_f/\Lambda^2_\mathrm{QCD}]]}{\ln[\mu^2_f/\Lambda^2_\mathrm{QCD}]} \right]}\approx2.27.
\end{equation}
Here, we have chosen $(\mu_i,\mu_f)=(0.6,2)$ GeV and $\Lambda_\mathrm{QCD}=0.2$ GeV as a trial, and $(\beta_0,\beta_1)=(11-2N_f/3,102-38N_f/2)/(4\pi)^2$. In Ref.~\cite{Bali:2012jv}, it turns out that $\mathcal{R}(1\,\mathrm{GeV}\to2\,\mathrm{GeV})\approx1.96$ from the four-loop order RG evolutions. On the contrary, we have $\mathcal{R}(1\,\mathrm{GeV}\to2\,\mathrm{GeV})\approx1.59$ at the two-loop level as in Eq.~(\ref{eq:RG}). Thus, we observe about $20\%$ difference in $\mathcal{R}$ between the two-loop and four-loop evolutions. For convenience, we define and compute the following quantities for running up the renormalization scale $(0.6\to2)$ GeV via Eq.~(\ref{eq:RG}):
\begin{equation}
\label{eq:RMCQC}
\mathcal{R}_\mathrm{MC}=\mathcal{R}^{\frac{\gamma_\mathrm{MC}}{b}}(0.6\,\mathrm{GeV}\to2\,\mathrm{GeV})\approx0.95,\,\,\,\,
\mathcal{R}_\mathrm{QC}=\mathcal{R}^{\frac{\gamma_\mathrm{QC}}{b}}(0.6\,\mathrm{GeV}\to2\,\mathrm{GeV})\approx1.40.
\end{equation}
Combining Eq.~(\ref{eq:RMCQC}) and Eq.~(\ref{eq:RGG}), one is led to
\begin{equation}
\label{eq:GFDF}
\langle\bar{q}\sigma\cdot G q\rangle_{2\,\mathrm{GeV}}=\mathcal{R}_\mathrm{MC}\,
\langle\bar{q}\sigma\cdot G q\rangle_{0.6\,\mathrm{GeV}},\,\,\,\,
\langle\bar{q} q\rangle_{2\,\mathrm{GeV}}=\mathcal{R}_\mathrm{QC}\,
\langle\bar{q}q\rangle_{0.6\,\mathrm{GeV}},
\end{equation}
As understood by Eq.~(\ref{eq:GFDF}), MC is insensitive to the RG evolution, whereas QC gives about $40\%$ increase for the larger renormalization scale. We also note that $\mathcal{R}_\mathrm{QC}=1.40$ is compatible $\mathcal{R}_\mathrm{QC}=1.32$ for the four-loop RG evolution for $\mu=(1\to2)$ GeV, given in Ref.~\cite{Bali:2012jv}, showing only a few percent difference. Here is one caveat: Although the values of $\mathcal{R}_\mathrm{MC,QC}$ in Eq.~(\ref{eq:RMCQC}) can be changed at finite $T$, we assume the changes to be small, since the $T$-dependence in $\mu_{i,f}$ is largely canceled as in Eq.~(\ref{eq:RG}), so that we use them in Eq.~(\ref{eq:RMCQC}) for whole $T$ region for the RG evolutions for the condensates. 

\begin{figure}[t]
\begin{tabular}{cc}
\includegraphics[width=8.5cm]{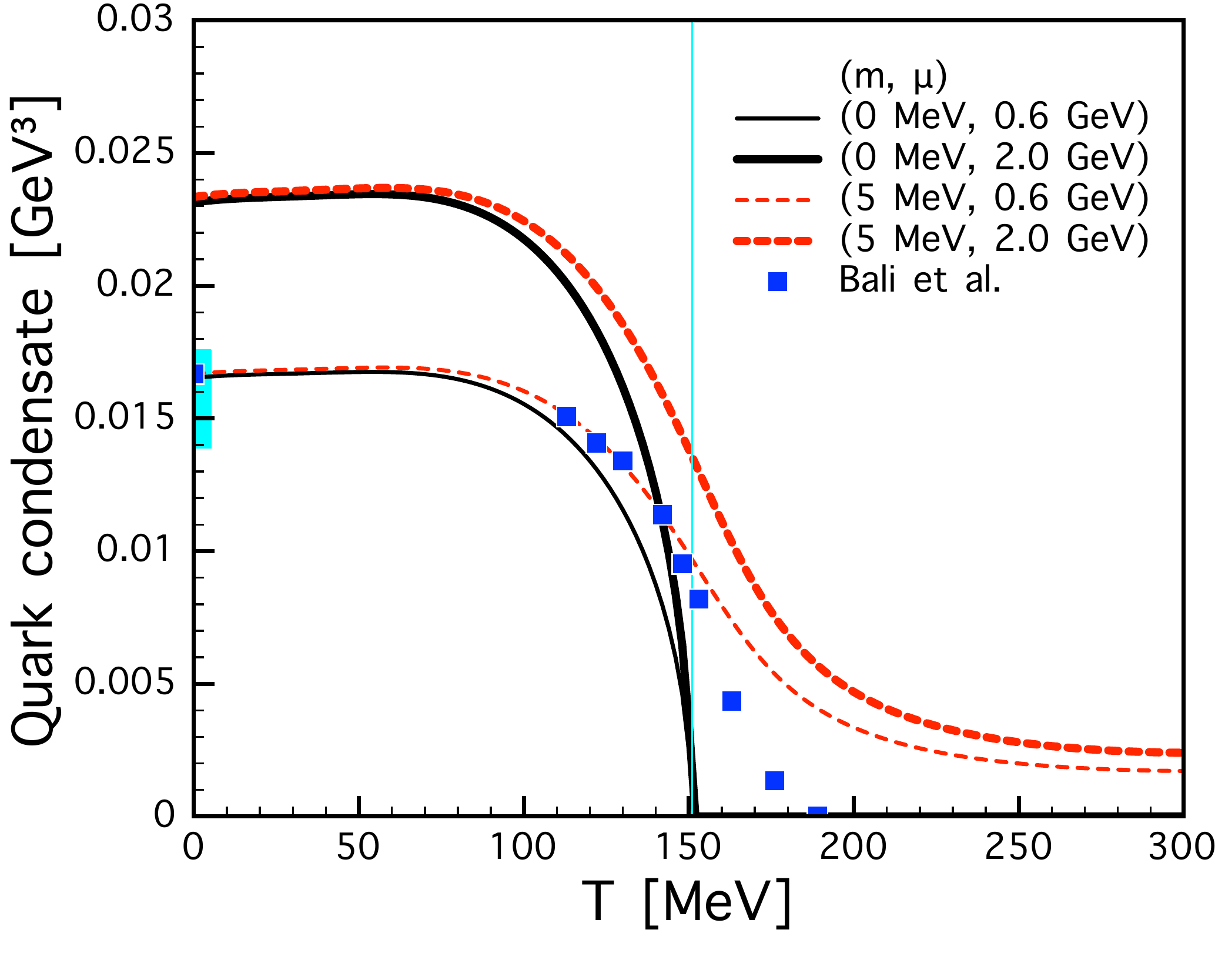}
\includegraphics[width=8.5cm]{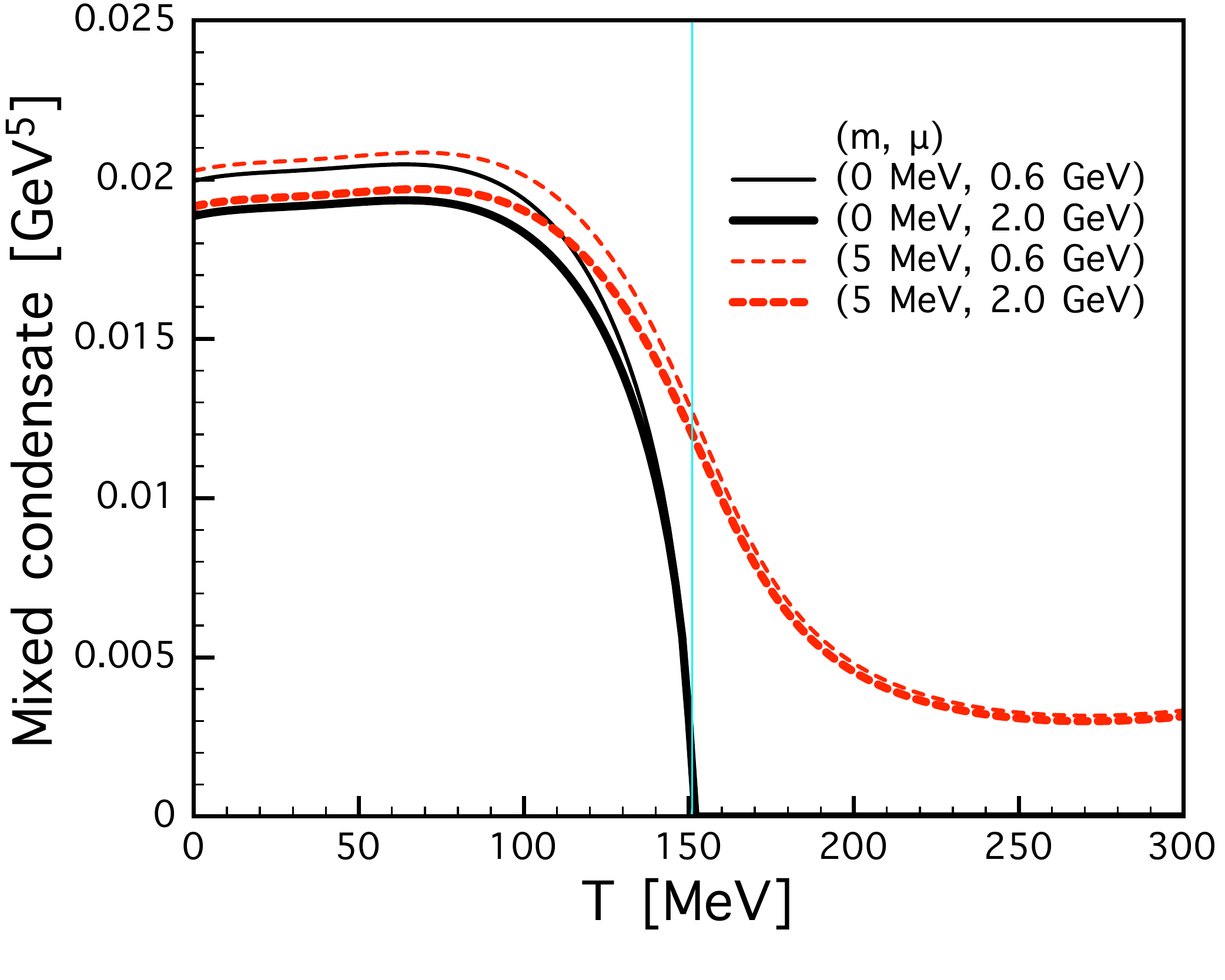}
\end{tabular}
\caption{(Color online) Left: Quark condensate (QC), $-\langle\bar{q}q\rangle$ [GeV$^3$] as functions of $T$ for different $m$ and $\mu$ values given in ($m,\mu$): (0 MeV, 0.6 GeV) (thin-solid), (0 MeV, 2.0 GeV) (thick-solid), (5 MeV, 0.6 GeV) (thin-dash), and (5 MeV, 2.0 GeV) (thick-dash). The vertical line indicates the chiral phase transition temperature $T_0\approx151$ MeV, whereas the shaded area denote the phenomenologically allowed region for QC in the chiral limit at $T=0$, i.e. $\langle\bar{q}q\rangle=-(240\sim260\,\mathrm{MeV})^3$. Right: Quark-gluon mixed condensate (MC), $-\langle\bar{q}\sigma\cdot G q\rangle$  [GeV$^5$], presented in the same manner with the left panel.}       
\label{FIG45}
\end{figure}

In the left panel of Figure~\ref{FIG45}, we show the numerical results for QC as functions of $T$ for $m=0$ (solid) and $m=5$ MeV (dash). The thin and thick lines stand for QC at the different renormalization scales $\mu=0.6$ GeV and $2.0$ GeV, respectively. For all the cases, we observe correct chiral restoration patterns, depending on the current-quark mass, as discussed above. The shaded area at $T=0$ denotes phenomenologically accepted range for QC: $\langle\bar{q}q\rangle=-(240\sim260\,\mathrm{MeV})^3$, and the present numerical results at $T=0$ for $\mu=0.6$ GeV match well with them as shown there. In Ref.~\cite{Bali:2012zg}, a LQCD simulation with the stout smeared staggered fermions was performed for QC. The LQCD data were extrapolated to the continuum limit. In their work, the transition $T$ was given by $T_0=158$ MeV, which is about $4\%$ larger than ours $T_0=151$ MeV. The LQCD data are also represented in the left panel of Figure~\ref{FIG45} with the solid squares. We normalize the LQCD data to be $(255\,\mathrm{MeV})^3$ at $T=0$ to match with the present numerical result for QC for $(m,\mu)=(5\,\mathrm{MeV},0.6\,\mathrm{GeV})$. As shown in the left panel of Figure~\ref{FIG45}, the LQCD data are well comparable with the present result below $T_0$, and the deviation increases as $T$ grows. Note that our results remain finite even beyond $T=300$ MeV, whereas the LQCD data becomes almost zero at $T\approx190$ MeV. As the renormalization scale evolved to $\mu=2.0$ GeV, the strength of QC increases by a factor $1.40$ as given in Eq.~(\ref{eq:RMCQC}). 

In the right panel of Figure~\ref{FIG45}, we show the numerical results for MC as a function of $T$ in the same manner with the left panel. At $T=0$, we have $-\langle\bar{q}\sigma\cdot G q\rangle^{1/5}=457\,(459)$ MeV for $m=0\,(5)$ MeV. These values are well consistent with those from the effective QCD-like models~\cite{Polyakov:1996kh,Nam:2006ng}: $-\langle\bar{q}\sigma\cdot G q\rangle^{1/5}=(481\sim484)$ MeV. The main source for this difference $\sim20\%$ in the MC values comes from the simplification from Eq.~(\ref{eq:MCB}) to Eq.~(\ref{eq:MCBT}), i.e. ignoring $(k_4,p_4)$ in the numerator of the integrand in Eq.~(\ref{eq:MCB}). The $T$ dependence of the MC curves is very similar to that computed by the global color-symmetry model~\cite{Zhang:2004xg}, whose $T_0$ was estimated to be about $170$ MeV for $\mu_q=0$: MC increases slightly below $T_0$, then goes to zero with respect to $T$.

\begin{figure}[t]
\includegraphics[width=8.5cm]{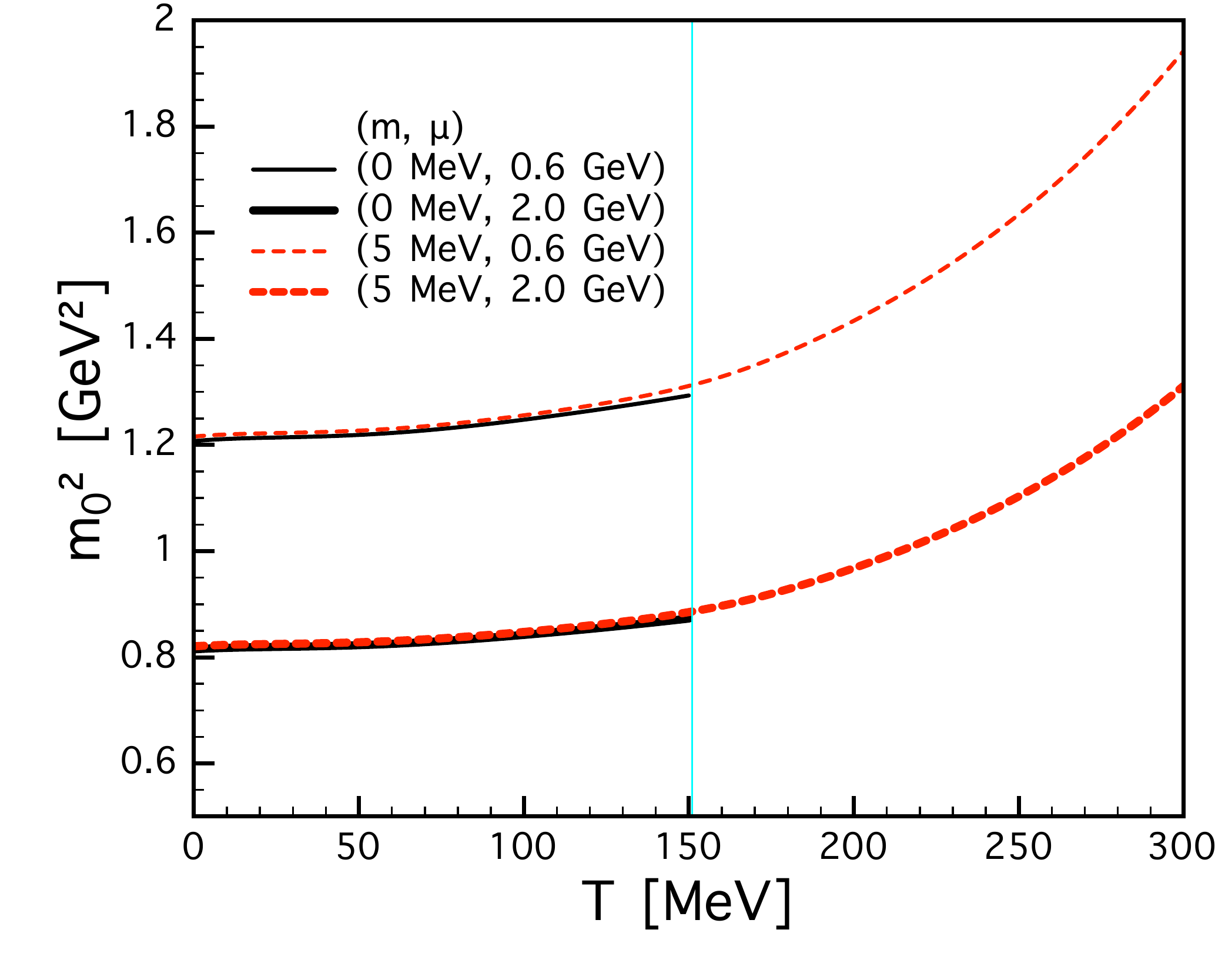}
\caption{(Color online) Ratio of the quark and mixed condensates $m^2_0$ in Eq.~(\ref{eq:QDEF}) as functions of $T$ for different $m$ and $\mu$ values given in ($m,\mu$): (0 MeV, 0.6 GeV) (solid), (0 MeV, 2.0 GeV) (dot), (5 MeV, 0.6 GeV) (dash), and (5 MeV, 2.0 GeV) (long-dash). The vertical line indicates the chiral transition temperature $T_0\approx151$ MeV.}       
\label{FIG6}
\end{figure}

The numerical results for the ratio of MC and QC are given in Figure~\ref{FIG6} in the same manner with Figure~\ref{FIG45}. At $T=0$, we have $m^2_0=1.12\,(1.22)\,\mathrm{GeV}^2$ for $m=0\,(5)$ MeV for $\mu=0.6$ GeV. For higher $\mu=2.0$ GeV, the ratios decrease slightly by a factor $0.95$ as in Eq.~(\ref{eq:RMCQC}). Again, these values are smaller than those at $\mu=0.6$ GeV in Refs.~\cite{Polyakov:1996kh,Nam:2006ng} by about $(10\sim20)\%$, due to the same reason for MC as discussed above. From the QCDSR calculations,  it was proposed that $m^2_0=(0.8\pm0.2)\,\mathrm{GeV}^2$ at $\mu=0.5$ GeV from a phenomenological point of view, i.e. QCDSR stability. From the global color-symmetry model~\cite{Zhang:2004xg}, the ratio was estimated to be $1.90\,\mathrm{GeV}^2$ from a simple confined Dyson-Schwinger method which is about $(30\sim40)\%$ larger than ours. Much larger value for $m^2_0$ was estimated from the SU($3_c$) KS-fermion LQCD simulation at the quenched level as $m^2_0=2.5\,\mathrm{GeV}^2$ at $\mu=2$ GeV, whereas ours is $0.83\,\mathrm{GeV}^2$ for $(m,\mu)=(5\,\mathrm{MeV},2\,\mathrm{GeV})$. As shown in the figure, the ratio curves are slightly increasing but almost flat below $T_0$ for all the cases. Taking  into account a simple linear parametrization for $m^2_0$ below $T_0$ at $\mu=0.6$ GeV, we have
\begin{equation}
\label{eq:PARAMSQR}
m^2_0(T)/m^2_0(0)\approx 0.07\,T/T_0+1\,\,\,\,\mathrm{for}\,\,\,\,T\lesssim T_0.
\end{equation}
This almost-flat  behavior is well consistent with those from the SU($3_c$) LQCD simulation with the KS fermion~\cite{Doi:2004xe}, although their $T_0$ is much higher than ours, i.e. $T_0\approx280$ MeV, due to the quenched simulation and the heavier current-quark mass $m=(20\sim50)$ MeV. In the global color-symmetry model~\cite{Zhang:2004xg}, the ratio $m^2_0$ does not depend on $T$ as well as $\mu_q$, and this observation is in qualitatively agreement with ours below $T_0$. 

Beyond $T_0$, however, it turns out that our results for $m^2_0$ for $m=5$ MeV increase rapidly with respect to $T$. This tendency can be easily understood by seeing the curves for QC and MC in Figure~\ref{FIG45}: The curves for QC keep decreasing up to $T=300$ MeV, whereas those for MC are almost saturated beyond $T=250$ MeV. It must be interesting to verify this increasing behavior in future LQCD simulations with proper extrapolation to the physical pion mass. Similarly to Eq.~(\ref{eq:PARAMSQR}), we have the following parametrization approximately for $m^2_0(T)/m^2_0(0)$ for $\mu=0.6$ GeV beyond $T_0$:
\begin{equation}
\label{eq:PARAMSQR2}
m^2_0(T)/m^2_0(0)\approx 0.47\,T/T_0+0.6\,\,\,\,\mathrm{for}\,\,\,\,T\gtrsim T_0.
\end{equation}
As understood easily from Eqs.~(\ref{eq:PARAMSQR}) and (\ref{eq:PARAMSQR2}), $m^2_0$ increases rapidly after $T\approx T_0$, in comparison to that below $T_0$. We also note that the ratio $m^2_0$ is related to the expectation value of the transverse momentum for the twist-$2$ light-cone distribution amplitude for the pion:
\begin{equation}
\label{eq:TR}
\langle k^2_T\rangle_\pi
=\frac{5}{36}\frac{\langle\bar{q}\sigma\cdot G q\rangle}{\langle\bar{q}q\rangle}
=\frac{5m^2_0}{36}.
\end{equation}
At $T=0$, we have $\langle k^2_T\rangle_\pi\approx0.17\,\mathrm{GeV}^2$ for $m=(0,5)$ MeV for instance. This value is well consistent with those from the chiral quark and instanton models in Refs.~\cite{Nam:2006au,Nam:2006ng}: $\langle k^2_T\rangle_\pi\approx=(0.20\sim0.23)\,\mathrm{GeV}^2$. All the numerical results for QC, MC, and $m^2_0$ are summarized for some different $T$ values in Table~\ref{TABLE1}.

\begin{table}[b]
\begin{tabular}{c|ccccccccc}
&$T=0$ [MeV]&$25$&$50$&$75$&$100$&$125$&$150$&$175$&200\\
\hline
$-\langle\bar{q}q\rangle^{1/3}$ [MeV]&$255\,(255)$&$255\,(256)$&$255\,(257)$&$255\,(256)$&$249\,(252)$&$232\,(240)$&$135\,(214)$&$0\,(177)$&$0\,(150)$\\
$-\langle\bar{q}\sigma\cdot G q\rangle^{1/5}$ [MeV]&$457\,(459)$&$458\,(460)$&$459\,(461)$&$459\,(461)$&$454\,(458)$&$437\,(447)$&$317\,(419)$&$0\,(376)$&$0\,(344)$\\
$m^2_0$ [GeV$^2$]&$1.21\,(1.22)$&$1.21\,(1.22)$&$1.22\,(1.23)$&$1.23\,(1.24)$&$1.25\,(1.26)$&$1.27\,(1.28)$&$1.29\,(1.31)$
&$-\,(1.36)$&$-\,(1.43)$
\end{tabular}
\caption{Quark condensate (QC) $\langle\bar{q}q\rangle$, mixed condensate (MC) $\langle\bar{q}\sigma\cdot G q\rangle$, and their ratio $m^2_0$ at different $T$ for $m=0\,(5)$ MeV. The renormalization scale is chosen to be $\mu=0.6$ GeV for all the cases. To evolve these values to $\mu=2$ GeV, one multiplies $1.40^{1/3}=1.12$, $0.95^{1/5}=0.99$, and $0.68$ to QC, MC, and $m^2_0$, respectively, as given in Eq.~(\ref{eq:RMCQC}).}
\label{TABLE1}
\end{table}

\section{Summary and conclusion}
We have investigated the quark-gluon mixed condensate (MC) at finite temperature ($T$), employing the $T$-modified liquid-instanton model (LIM-$T$). In the present theoretical framework, one can write the quark-gluon Yukawa interaction in terms of the effective nonlocal four-quark vertex. Using it, we calculated MC at a low renormalization $\mu=0.6$ GeV, then performed the RG evolution up to $\mu=2$ GeV. Since we were interested in the $T$ dependence of MC, we obtained the $T$-dependent instanton parameters, $\bar{\rho}(T)$ and $\bar{R}(T)$ from the trivial-holonomy caloron solution. We also made use of the effective thermodynamic potential and fermionic Matsubara formula to determine $M_0(T)$ and to compute the relevant condensates. Numerical results for MC, quark condensate (QC), and their ratio $m^2_0\equiv\mathrm{MC/QC}$ were presented with discussions. Important observations in the present work are listed below:
\begin{itemize}
\item The $T$-modified instanton parameters show partial partial chiral restoration behaviors as $T$ increases: The average instanton size $\bar{\rho}(T)\approx1/\mu$ and instanton number density $N/V(T)$ slowly decrease with respect to $T$, indicating the reduction of the nonperturbative effect, which is given by the nontrivial interaction of the quarks and (anti)instantons via the quark zero mode. Note that this $T$ dependence of the relevant model parameters is a peculiar feature of the present model.  
\item By solving the saddle-point equation for the effective thermodynamic potential, we obtain the chiral order parameter $M_0$, i.e. constituent-quark mass, and the chiral phase transition $T$ is given by $T_0\approx151$ MeV for $m=(0,5)$. The chiral phase transitions are second-order and crossover for $m=0$ and $5$ MeV, respectively, satisfying the universal chiral restoration patterns.
\item Numerical results for QC are well compatible with the phenomenological values and the LQCD simulation data below $T_0$. However, we observe sizable deviation, in comparison to the LQCD data for $m\ne0$, beyond $T_0$. As $T$ grows beyond $T_0$, QC keeps decreasing but finite up to $T=300$ MeV. By running up the scale $\mu=(0.6\to2.0)$ GeV,  we had a multiplicable factor $1.40$ from the two-loop RG evolution. 
\item Those for MC exhibit similar curve shapes to that of QC as functions of $T$. It turned out that the MC curves increase slightly up to $T\approx100$ MeV, then start to decrease. This tendency is found to be similar to that from the global color-symmetry model.  As $m=5$ MeV, MC curves are almost saturated beyond a certain value in the vicinity of $T\gtrsim220$. We had a RG-evolution factor $0.95$ for $\mu=(0.6\to2.0)$ GeV.
\item The ratios $m^2_0$ are given numerically by $m^2_0=1.21\,(1.22)\,\mathrm{GeV}^2$ for $m=(0,5)$ MeV for vacuum. These values are well compatible with those from  other effective models and QCDSR estimation. It turned out that $m^2_0$ is almost flat  with the slope $\sim0.07$ for $T\lesssim T_0$ as a function of $T/T_0$, being consistent with the (quenched) LQCD simulation. Above $T_0$, the curve of $m^2_0$ increased rapidly, depending on the different $T$ dependence between QC and MC. We suggest an approximated linear parametrization for $m^2_0$ as:
\begin{equation}
\label{eq:PARAMSQR3}
m^2_0(T)/m^2_0(0)\approx (0.07,0.47)\,T/T_0+(1,0.6)\,\,\,\,\mathrm{for}\,\,\,\,(T\lesssim T_0,T\gtrsim T_0),
\end{equation}
\end{itemize}

The present results for MC as a function of $T$ will be a useful guide for future LQCD simulations and various model calculations. Note that the effects of the external strong electromagnetic field, which is expected to be created  from the peripheral (non-central) HIC, has been widely discussed~\cite{Abelev:2008ab,Fukushima:2008xe,Tuchin:2013ie}. In addition, the distribution of the gluons in the quark-gluon plasma (QGP) is an important input for the hydrodynamic studies for HIC~\cite{Ozonder:2012vw}. It is worth mentioning that, using the effective quark-gluon vertex as in the present work, one can explore the effects of the external EM field, which interacts with the gluon fields in terms of the quark ones. Related works are under progress and appear elsewhere.
\section*{Acknowledgments}
S.i.N. is grateful to G.~Endr\"odi (Regensburg) for fruitful comments. The numerical calculations were partially performed via the computing server ABACUS2 at KIAS.   

\end{document}